# Mean surface shape of a human placenta.

M. Yampolsky, O. Shlakhter, C. M. Salafia, D. Haas

**Abstract.**

**Goal:** The chorionic plate (or "fetal surface") of the human placenta is drawn as round, with the umbilical cord inserted roughly at the center, but variability of this shape is common. The average shape of the chorionic plate has never been established. The goal of this work is to measure the average shape in a birth cohort.

**Materials and Methods:** 1227 placentas collected from a prospective cohort had digital photographs taken of the chorionic plate after trimming of the membranes from the margin. A trained observer marked the perimeter of the fetal surface at intervals of between 1cm and 2cm. 1024 of the placentas were delivered at term. Of these, 44 cases were excluded because the measurement of chorionic plate landmarks was not possible. This left N=980 cases for analysis, 95.7% of all placentas delivered at term. Mathematical methods to estimate the average chorionic plate shape included calculating the *median shape,* and the *mean perimeter.* Mean values of the fetal surface area and the radial distance from the edge to the umbilical insertion point were also calculated.

**Results:** The average chorionic plate is round and centered about the umbilical cord insertion. The median shape is round, with the average radius from the umbilical insertion point equal to R=9.066 cm (range 8.76 − 9.46 cm, mean square deviation 2%). The mean perimeter method gives R=9.01cm (with range 8.71 − 9.27 cm, mean square deviation 1.9%). The mean chorionic plate area is calculated to be equal to 285.58 cm$^2$, corresponding to R=9.53 cm.

**Conclusions:** The average fetal surface shape for a placenta delivered at term is a disk around the umbilical insertion point. Different approaches to calculating the average agree on this and give similar estimates for the radius of the disk of about 9cm. Both the average radius from the umbilical insertion point, and the surface area of an individual placenta can serve as reliable measurements of the deviation from the mean shape.

**Introduction**

The chorionic plate (or "fetal surface") of the human placenta is typically drawn as round, with the umbilical cord inserted roughly at the center **[1]**. In clinical practice, the shape of the chorionic disk is rarely truly circular; its shape commonly varies, from round to oval, bi- or multi- lobate, or regularly irregular. Placental shape is thought to be determined by where it is implanted in the uterus, regional variations in the decidua (that may determine areas of atrophy), variations in maternal vascular supply (with placental infarcts resulting in altered shape) and perhaps even the "manner of its original implantation" **[1]**. Benirschke states that "the mechanism by which [they] develop is unclear", although "much supports the notion of a secondary conversion from more normal placentation" **[1]**.

In a recent paper **[2],** we have related this variability in chorionic plate shape to the structure of the underlying placental vascular tree. We could reliably produce multilobate and regularly irregular shape deviations from a typical round chorionic plate shape by changing a model parameter that affected the density of fractal growth of the chorionic villous tree. In view of the dependence on chorionic plate shape on the uniformity of fractal growth of the chorionic villous tree , it is particularly important to understand how typical the round chorionic plate shape is, and to obtain reliable statistics of chorionic plate shape. In this paper we use several different techniques to describe an *average* chorionic plate shape for placentas delivered at term. All our findings agree that the average chorionic plate shape is a circle around the umbilical insertion point. We develop several specific measurements of the deviation from the average which are both practical to implement, and statistically reliable.

**Materials and methods.**

*Placental Cohort.*

The *Pregnancy, Infection, and Nutrition Study* is a cohort study of pregnant women recruited at mid pregnancy from an academic health center in central North Carolina. Our study population and recruitment techniques are described in detail elsewhere **[3]**. Beginning in March 2002, all women recruited into the *Pregnancy, Infection, and Nutrition Study* were requested to consent to a detailed placental examination. As of October 1, 2004, 94.6 percent of women consented to such examination. Of those women who consented, 87 percent had placentas collected and photographed for image analysis. Of the 1,227 consecutive placentas collected, 203 were delivered pre-term and were excluded. Further 44 cases were excluded because the measurement of chorionic plate landmarks was not possible. This left N=980 cases for analysis.

Placental gross examinations, histology review, and image analyses were performed at *EarlyPath Clinical and Research Diagnostics*, a New York State-licensed histopathology facility under the direct supervision of Dr. Salafia. The institutional review board from the University of North Carolina at Chapel Hill approved this protocol.

The fetal surface of the placenta was wiped dry and placed on a clean surface after which the extraplacental membranes and umbilical cord were trimmed from the placenta. The fetal surface was photographed with the Lab ID number and 3 cm. of a plastic ruler in the field of view using a standard high-resolution digital camera (minimum image size 2.3 megapixels). A trained observer (D.H.) captured series of *x,y* coordinates that marked the site of the umbilical cord insertion, the perimeter of the fetal surface, and the "vascular end points", the sites at which the chorionic vessels disappeared from the fetal surface. The perimeter coordinates were captured at intervals of between 1cm and 2cm, and more coordinates were captured if it appeared essential to accurately capturing the shape of the fetal surface.

*Software.*

Numerical calculations were carried out using *mean-placenta*, a Unix-based, ANSI C, software package developed under the terms of the GNU General Public License as published by Free Software Foundation. For visualizations we have used *PovRay*: a freeware ray tracing program available for a variety of computer platforms; and Maplesoft Maple 9.0 Math & Engineering software.

**Calculating the median shape of the chorionic plate.**

A square grid of 500x500 pixels was superimposed on the images of the placentas in the sample, which were rescaled to the actual size. The size of one pixel was chosen to equal 0.1cm, so the total coverage of the grid was a square of 50x50 cm$^2$. The insertion point of the umbilical cord was placed at the center of the square, and the point on the placental perimeter closest to the rapture of the amniotic sac was placed on the negative vertical axis, for consistency.

For each of the placentas in the dataset, all the pixels inside of the placental perimeter, including those intersecting with the boundary, were marked. Thus, for each placenta, an approximate area covered by it was obtained as the union of the marked pixels. Let us introduce some notation. For the *n*-th placenta in the dataset, we will denote $S_n$ the region of the 50x50-square, which is covered by its surface. We will let $W_n$ denote the union of the marked pixels for this placenta. Note that the edge of $W_n$ is no more than the size of one pixel (0.1cm) removed from the edge of $S_n$.

To calculate the mean shape of the chorionic plate, for each pixel *p(x,y)*, *x=1…500, y=1…500* of the grid, we marked the total number *t(x,y)* of the placentas for which *p(x,y)* lies in $W_n$. The central pixels will thus be covered by all of the placentas, that is *t(x,y)≈N* (the total number of placentas in the dataset) for *x,y≈250*, and the peripheral ones will be covered by very few.

We let $W_{median}$ be the union of all pixels *p(x,y)* with *t(x,y)≥N/2*. It is thus the *median shape* of the chorionic plate in our sample. The median chorionic plate shape is round, as seen in Figure 1. We calculated the mean radius of a pixel on the boundary of $W_{median}$ from the umbilical insertion point. Its value is

*Mean Radius($W_{median}$)= 9.066cm.*

The range of radii of boundary pixels was from 8.76cm to 9.464cm, and the mean square deviation from 9.066cm was 0.181cm, or 2%.

We have also calculated the shapes $W_{40\%}$ and $W_{30\%}$ which correspond to the values of *t(x,y)* greater than *0.4N* and *0.3N,* respectively. They are also round (Figure 1). This indicates, that the deviations of a chorionic plate shape from the median have radial symmetry – they are *equally likely to happen in every direction* from the umbilical insertion point.

**Calculating the mean placental perimeter.**

To calculate the mean chorionic plate shape, the insertion point of the umbilical cord was again placed at the origin, and the point on the perimeter closest to the rapture of the amniotic sac was positioned on the negative vertical axis. The perimeters of the placentas in the dataset were rescaled to the real size. The points in the perimeter were then averaged inside a sector of 18º, thus obtaining 20 radial markers for each placenta. This procedure is illustrated in Figure 2. Each of the markers was then averaged over the whole dataset, thereby giving 20 mean placental radii from the umbilical insertion point spaced at 18º angular intervals. They were further averaged to obtain the mean average chorionic plate radius

$$R_{average}=9.01cm.$$

The 20 mean radii were within 3.3% of the average chorionic plate radius, with the maximal one equal to 9.27cm, and the minimal one equal to 8.71cm. The mean square deviation from the average chorionic plate radius was equal to 0.18cm or 1.9%. Thus, the mean perimeter shape is round, with a centrally inserted umbilical cord, as seen in Figure 3.

We note that this method of calculating an average chorionic plate perimeter is potentially biased towards making irregular-shaped chorionic plate appear more round. An extreme example of this would be a horseshoe-shaped placenta, with the insertion point of the umbilical cord in one of the ends of the horseshoe. After averaging the radii, it would appear as a round chorionic plate – the empty space in the middle would be ignored by this method. However, the result is in such an excellent agreement with the computation of the median chorionic plate shape, that apparently no significant bias has been introduced.

**Practical measurements: the average chorionic plate area and radius.**

Having confirmed that the mean shape of a chorionic plate is a round disk, we describe two simple and practical measurements which may be used to evaluate the deviation of the chorionic plate shape from a round one.

We have calculated the average area of the region covered by the chorionic plate shape as the arithmetic mean of areas of $W_n$

$$\text{Average Area} = \sum_{n=1,\ldots,N} \text{Area}(W_n)/N = 285.58 cm^2.$$

The standard deviation of the area from the average in our sample is 60.78cm$^2$ or 21.2%. While the standard deviation appears large, recall that for a disk of radius $R$, the change of radius by $\Delta R$ changes the area by approximately $2\pi \cdot \Delta R$. Thus even if our placentas were all round, a 21.2% variance of the area would correspond to less than 3.4% of radial variance.

A *circle* with area 285.58cm$^2$ has radius 9.53 in an excellent agreement with our previous findings. This is particularly impressive, as the overall variance of chorionic plate areas in our sample is quite large: between the minimal value 135.03cm$^2$ and the maximal value 649.44cm$^2$.

Note that $W_n$ is slightly larger than the true chorionic plate $S_n$, as the boundary pixels which only partially intersect $S_n$ are wholly included into $W_n$. Thus the *Average Area* computed by our method has a slightly larger value than the true average of the chorionic plate areas. However, the bias is not very significant. One way to estimate it is to assume that $S_n$ is round, in which case the radius would be increased by about one half of the size of a pixel, which in our case is 0.05cm.

As a second measurement, we have calculated the average radius $R_n$ of a chorionic plate in our dataset by averaging the 20 radial markers obtained as described above. The standard deviation of $R_n$ from the mean value $R_{average}=9.01$cm is 1.07cm, or 11.8%.

Thus both the chorionic plate area, and the outer radius from the umbilical insertion point are reliable measurements of the regularity of the chorionic plate shape. The 95% confidence intervals for chorionic plate area and chorionic plate radius are [281.77cm$^2$, 289.38 cm$^2$] and [8.94 cm, 9.07 cm] respectively.

**Conclusion.**

We have shown that the average shape of the chorionic plate is a disk around the umbilical insertion point. While this has been conventionally believed, the burgeoning ultrasonography literature that has linked various detectable markers of atypical placental shape to later significant fetal morbidity and/or mortality has made a deeper understanding of the genesis of placental shape imperative. In previous work **[2]**, we have shown that the fetal surface shape closely mirrors the pattern of chorionic vascular fractal growth.

The principal ultrasonographic markers linking altered placental shapes in the midtrimester with later poor outcomes are increased disk thickness and abnormally eccentric umbilical cord insertion. Abnormal placental shape at 19-23 weeks was defined by Toal and colleagues **[4]** principally in terms of abnormal thickness (thickness of > 4 cm, or 50% of length) and associated with higher odds of intrauterine fetal death, extreme preterm delivery, and fetal growth restriction. The study placental gross examination protocol did not record chorionic plate shape. We note that, in the placental vascular growth model developed in our work **[2],** deviations from a round to oval shape also created variations in the thickness of the fractal structure of the vasculature, consistent with the thickness variability common in irregularly shaped placentas.

Eccentricity of the umbilical cord insertion has also been proposed as a clinically useful marker of "placental insufficiency" **[5], [6].** These authors considered "the most serious forms of placental insufficiency" to be "characterized by small placentas with eccentric cords, termed 'chorionic regression'", and further speculated that such regression was likely caused by inadequate endovascular trophoblast function in the placental bed. They also concluded that the "default response of the placenta to uteroplacental vascular insufficiency is to increase placental angiogenesis… becoming thicker". Not surprisingly, given their proposed common basis in maternal uteroplacental vascular pathology, eccentric cord insertion has also been linked to abnormal placental shape in cases of severe IUGR with abnormal umbilical artery Doppler findings **[7]**.

In our previously published model for chorionic vascular fractal growth **[2]** a non-round chorionic plate shape could be produced by changes in the branching density of the vascular tree at different points in gestation. That parameter could represent alterations in the maternal vascular or immunological environments, changes in intrauterine growth factors, or other molecular mechanisms. The connection between variations in the fetal surface shape and the structure of the underlying vasculature is consistent with above quoted findings.

Several mathematical approaches developed in the present work yield similar estimates on the radius of the average round shape of the chorionic plate of about 9cm. We have demonstrated that both the outer radius and the surface area of the chorionic plate can be used to estimate the deviation from the mean shape, and have provided confidence limits for both quantities. These measurements can thus be useful clinical tools in the study of variability of the placental shape and its effect on the placental function.

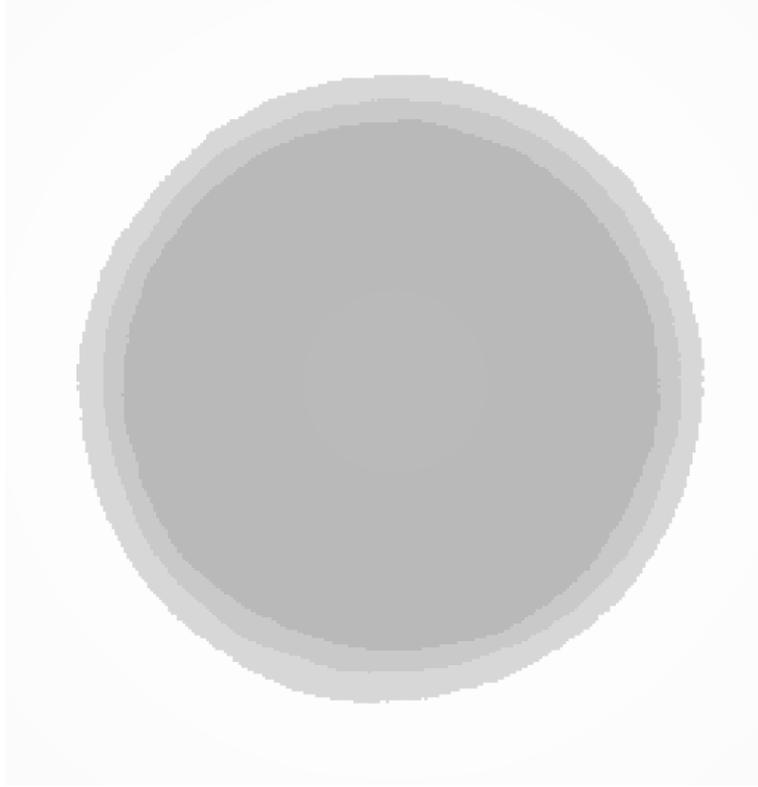

**Figure 1.** Median placental shape $W_{median}$ (smallest disk) and $W_{40\%}$, $W_{30\%}$ (two larger disks). The position of the insertion point of the umbilical cord is in the center of the picture.

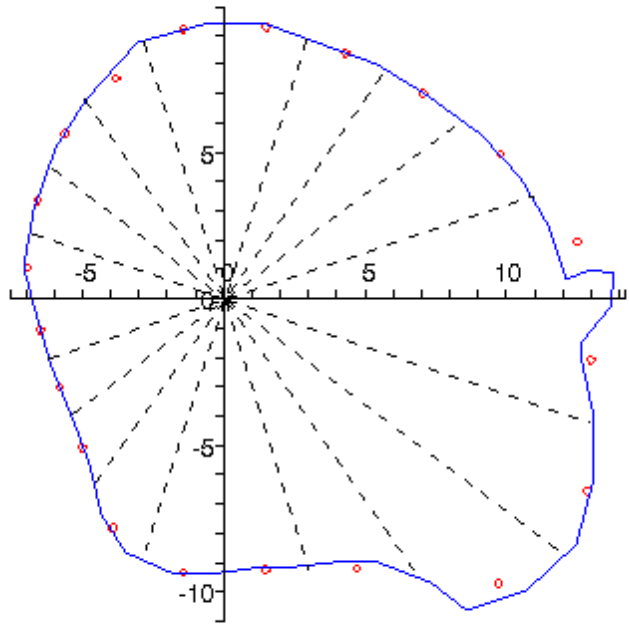

**Figure 2.** Calculation of the 20 radial markers for one of the placentas in the dataset. The solid contour marks the perimeter of the placental surface. The 20 radial sectors emanate from the umbilical insertion point. Each marker is indicated with a small circle. It averages the distance from the perimeter curve to the insertion point within the sector.

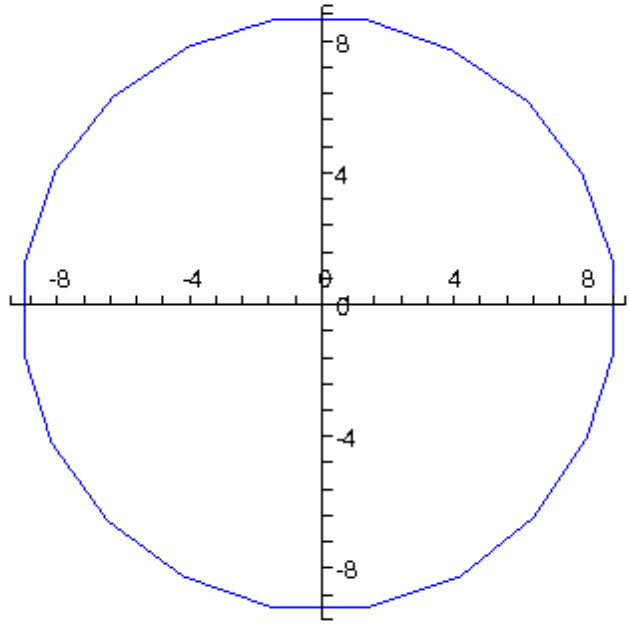

**Figure 3.** The positions of the 20 mean perimeter markers, connected to show the mean placental perimeter. The umbilical insertion point is at the origin.